\documentclass[twocolumn,showpacs,preprintnumbers,amsmath,amssymb,groupedaddress,pre]{revtex4}

\usepackage{graphicx}
\usepackage{dcolumn}
\usepackage{bm}
\usepackage{epsfig}

\def\e{\begin{equation}}
\def\f{\end{equation}}
\def\=#1{\overline{\overline #1}}

\def\-#1{{\bf #1}}
\def\.{\cdot}
\def\l#1{\label{eq:#1}}
\def\r#1{(\ref{eq:#1})}

\def\dyad#1{\overline{\overline {\bf{#1}} }}

\begin{document}

\title{Additional Boundary Conditions for Nonconnected Wire Media}

\author{M\'ario G. Silveirinha}
 \altaffiliation[]{Electronic address: mario.silveirinha@co.it.pt}
 \affiliation{University of Coimbra, Department of Electrical
  Engineering-Instituto de Telecomunica\c{c}\~{o}es, 3030 Coimbra, Portugal}

\date{\today}

\begin{abstract}
Following our recent work [New J. Phys. 10, 053011, (2008)], here we
demonstrate that due to strong nonlocal effects additional boundary
conditions are essential to characterize the reflection of
electromagnetic waves by nonconnected wire arrays using
homogenization methods. Based on simple physical considerations, we
derive the additional boundary conditions for the case where the
wire medium is adjacent either to a dielectric or to a conducting
material, and demonstrate that in the lossless case such boundary
conditions ensure the conservation of the power flow. It is shown
that the number of additional boundary conditions is related to the
number of metallic wires in a unit cell.  We illustrate the
application of the novel boundary conditions to several
configurations with practical interest.
\end{abstract}

 \maketitle

\section{Introduction}
Artificial materials formed by long metallic wires have attracted
significant attention in recent years \cite{PendryPlasmons,
SmithDNG, Pokrovsky_1, WMPRB, IgorWM2D, IgorPRE, SWIWM, Shin,
ShvetsPendry, PekkaAPL, 1meter, ZhangScience,  ZhangOE, MarioEVL}.
In some circumstances such wire media may be characterized by a
plasmonic-type electric response, which potentially may enable
interesting phenomena to occur such as superlensing
\cite{Silv3DWMIR}, negative refraction \cite{ZhangScience, ZhangOE},
or the excitation of strongly localized electromagnetic modes.
However, it has been known for some time that in general wire media
may fail to completely mimic the properties of a local plasma, and
may have a spatially dispersive response \cite{WMPRB,
Silv_Nonlocalrods, Pokrovsky_2, Silv_MTT_3DWires, Constantin_WM,
IgorPRE, ShvetsWM, Shapiro, NonlocalENZ}, i.e. the electric
displacement vector in a given point of space cannot be written
exclusively in terms of the macroscopic field at the considered
point, but, on the contrary, may depend on the distribution of the
electric field in a neighborhood that encompasses many unit cells.
Such property is a consequence of the fact that the metallic wires
are spanned over many unit cells, and consequently the polarization
acquired by the metallic inclusions in a given unit cell, (which is
roughly proportional to the current along the wires), may depend
significantly on the electric field outside the considered cell.
Recently, it was suggested that the spatial dispersion effects may
be tamed either by attaching conducting structures on the wires or
by coating the wires with a magnetic material \cite{PendryDemi}. In
Ref. \cite{Silv3DWMIR} it was also demonstrated that at infrared and
optical frequencies the nonlocal effects may be significantly
weakened when the radius of the metallic rods is comparable to (or
smaller than) the skin depth of the metal.

Even though in some applications spatial dispersion effects may be
regarded as undesirable, in other circumstances they may open new
possibilities. For example, in part due to nonlocal effects, an
array of parallel wires may behave as a material with extreme
optical anisotropy and this may enable the transport and
manipulation of the near-field \cite{SWIWM, WMIR, WMLimitRes,
ShvetsPendry, PekkaAPL, 1meter}. Moreover, in a recent series of
works it was shown that the spatially dispersive properties of a
crossed wire mesh may enable the realization of materials with an
extreme index of refraction \cite{MarioEVL}, the realization of
ultra-subwavelength waveguides \cite{MarioEVL, EVLexp}, superlensing
\cite{XwiresLens}, broadband all-angle negative refraction
\cite{XWiresNegRef}, and low-loss broadband anomalous dispersion
\cite{XwiresRainbow}. Some of these effects (e.g. the low-loss
anomalous spectral dispersion) are specific of spatially dispersive
materials, and cannot be observed in local materials. Due to these
and other potential applications, the accurate characterization and
modeling of nonlocal materials gains increasing importance.

One of the peculiarities of nonlocal materials is that they may
support additional waves, i.e. for a fixed direction of propagation
and a fixed frequency, the number of plane waves supported by the
material may be greater than two, differently from conventional
local materials. For example, some of the plane waves may have
identical polarization (i.e. the same orientation for the electric
field) and be associated with different wave vectors. It is well
known that such property implies that the classical boundary
conditions, which impose the continuity of the tangential components
of the electric and magnetic fields, may be insufficient to solve a
scattering problem at an interface between a spatially dispersive
material and another material, due to the degrees of freedom
associated with the ``new'' waves. A possible strategy to overcome
this problem is the introduction of additional boundary conditions
(ABC) \cite{Mahan, Melnyk, Pekar, Davis_ABC}. In Refs.
\cite{Henneberger, IgorWM2D} alternative ``ABC-free'' strategies are
described.

There is no general theory for the derivation of ABCs. In fact, the
ABCs depend on the very specific microstructure of the material, and
should describe the dynamics of the internal variables of the
nonlocal material, so that the degrees of freedom are removed. In
recent works the ABC approach was successfully used to model the
reflection and refraction of electromagnetic waves at an interface
between a wire medium formed by parallel wires and a conventional
(dielectric or conducting) material \cite{MarioABC, ABCtilted}.

The objective of this paper is to generalize the theory of our
previous works \cite{MarioABC, ABCtilted} to the case of
nonconnected ``double'' and ``triple'' wire media, and demonstrate
that the introduction of suitable ABCs enables the accurate modeling
of these structures using homogenization techniques. As mentioned
before, such materials may have interesting potentials in several
problems \cite{MarioEVL, EVLexp, XwiresLens, XWiresNegRef,
XwiresRainbow}, which justifies, besides obvious theoretical
motivations, the present study. It should be mentioned that it is
not trivial to extend the theory of Refs. \cite{MarioABC, ABCtilted}
to the case of more complex metamaterial topologies. For example, it
will be shown here that in general at an interface between a crossed
wire mesh and a regular dielectric two distinct ABCs are required,
whereas for an array of parallel wires one single ABC is sufficient
to characterize the interface effects \cite{MarioABC, ABCtilted}.

This paper is organized as follows. In section \ref{ABC}, we derive
the ABCs. In section \ref{SectPowerFlow} it is demonstrated that in
the lossless case the proposed ABCs ensure the conservation of the
power flow through a wire medium slab. A formula for the Poynting
vector in the spatially dispersive material is derived. Then, in
sections \ref{sectNum1}-\ref{SectXOZ}, we illustrate the application
of the ABCs in several scenarios of interest, and demonstrate their
accuracy using full wave numerical simulations. In particular, we
discuss the potentials of a wire mesh in the realization of
ultra-subwavelength waveguides and the emergence of negative
refraction. Finally, in section \ref{SectConclusion} the conclusions
are drawn. It is assumed that the fields are monochromatic and have
the time dependence $e^{ - i\omega t}$.

In this work, we use several different notations for the field
entities. For the convenience of the reader, the meaning of these
notations is summarized below for the electric field case:
\begin{itemize}
\item
${\bf{e}}$ - microscopic electric field. This is the exact solution
of the Maxwell's equations, taking into account the exact
microstructure of the material.
\item
${\bf{E}}$ - ``bulk'' electric field. This is the electric field
that results from the homogenization of the bulk material.
Specifically, the macroscopic field is obtained by averaging the
microscopic field over a volumetric region within which the
structure is invariant to translations along three independent
directions of space.
\item
${\bf{E}}_{{\rm{av,T}}}$ - transverse averaged electric field. This
definition requires that the considered structure is periodic along
two directions of space (assumed parallel to the $xoy$ plane).
However, the periodicity along $z$ is not required. The transverse
averaged field is obtained by averaging the microscopic field
exclusively along the directions parallel to the $xoy$ plane, as
detailed in section \ref{ABC}.
\end{itemize}

\section{Derivation of the Additional Boundary Conditions \label{ABC}}

Here, we generalize the ideas of \cite{MarioABC, ABCtilted} and
derive additional boundary conditions that enable the
characterization of the internal variables responsible for the
spatial dispersion effects in nonconnected wire media.

\begin{figure}[th] \centering
\epsfig{file=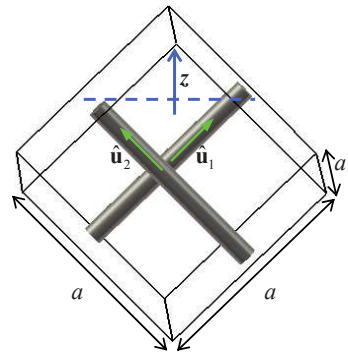, width=5.0cm} \caption{(Color online)
Geometry of the unit cell of the ``bulk'' double wire medium
($N=2$). The cubic unit cell contains two nonconnected metallic
wires. The dashed line represents an hypothetical cut of the bulk
material with a plane normal to the z-direction.} \label{cell}
\end{figure}

The material is formed by $N$ mutually orthogonal and nonconnected
sets of wires, oriented along the directions of space ${\bf{\hat
u}}_n$ ($n=1..N$) \cite{Silv_MTT_3DWires}, where $N=1,2,3$ is the
number of components of the wire medium. Each wire array is arranged
in a square lattice with lattice constant $a$. The radius of the
wires is $r_w$. For $N=1$ the material consists of a single array of
parallel wires, and for $N=2$ ($N=3$) it is formed by a double
(triple) wire array. The distance between adjacent orthogonal wires
is $a/2$. It is assumed that the wires are good conductors, i.e. the
radius of the wires is much larger than the skin depth of metal at
the frequency of operation. For reasons that will be clear shortly,
in general the unit vectors ${\bf{\hat u}}_n$ are not oriented along
the coordinate axes $x,y,z$.

As an example, the unit cell of the bulk double wire medium (crossed
wire mesh) is represented in Fig. \ref{cell}. The unbounded material
is formed by the periodic repetition of the primitive cell. Since
the wires intersect the boundaries of the unit cell, it should be
clear that such construction effectively yields infinitely long
wires.

Let us suppose that the bulk material is sliced to form a planar
interface with another adjacent material (a regular dielectric or
metal; see Fig. \ref{Compar}c). This operation breaks the
translational symmetry of the system along the direction normal to
the interface, which without loss of generality is assumed to be the
$z$-axis. In particular, the wires lying in cells of the wire medium
adjacent to the interface are sliced as well (except, possibly, if
some of the wires are parallel to the interface). However, the
periodicity of the system in the transverse plane ($xoy$-plane) is
preserved.

We use the transverse-average (TA) field approach introduced in
\cite{Silv_TAfield}, in order to describe the electromagnetic wave
propagation using homogenization methods. The TA-fields are obtained
by averaging the ``microscopic'' fields over the directions parallel
to the interface ($x$ and $y$ directions), i.e. over the directions
along which the system has translational symmetry. It is assumed
that the microscopic fields have the Floquet-Bloch property in the
$xoy$ plane, being characterized by the transverse wave vector
${\bf{k}}_{||} = \left( {k_x ,k_y ,0} \right)$. This situation
occurs, for example, when the wire medium slab is illuminated by a
plane wave, being in such circumstances ${\bf{k}}_{||}$ determined
by the direction of incidence and by the frequency of operation.

Following Ref. \cite{Silv_TAfield}, the TA-electric and induction
fields are defined as:
\begin{eqnarray}
{\bf{E}}_{{\rm{av,T}}} \left( z \right) = \frac{1}{{A_{cell}
}}\int\limits_{\Omega _T}
{{\bf{e}}\left( {\bf{r}} \right)e^{-i{\bf{k}}_{||} {\bf{.r}}} dxdy} \nonumber \\
{\bf{B}}_{{\rm{av,T}}} \left( z \right) = \frac{1}{{A_{cell}
}}\int\limits_{\Omega _T} {{\bf{b}}\left( {\bf{r}}
\right)e^{-i{\bf{k}}_{||} {\bf{.r}}} dxdy} \l{TAdef}
\end{eqnarray}
where ${\bf{e}}$ and ${\bf{b}}$ are the microscopic electric and
induction fields, $\Omega _T$ represents the transverse unit cell in
each $z=const.$ plane, and $A_{cell}$ is the area of $\Omega _T$. As
demonstrated in Ref. \cite{Silv_TAfield}, the TA-fields verify the
the differential system (assuming the time convention $e^{-i\omega
t}$):
\begin{eqnarray}
\left( { i{\bf{k}}_{||}  + \frac{d}{{dz}}{\bf{\hat u}}_z } \right)
\times \frac{{{\bf{B}}_{{\rm{av,T}}} }}{{\mu _0 }} &=& -i\omega
\varepsilon _0 \varepsilon_h {\bf{E}}_{{\rm{av,T}}} + {\bf{J}}_{d,{\rm{av}}} \nonumber \\
\left( { i{\bf{k}}_{||}  + \frac{d}{{dz}}{\bf{\hat u}}_z } \right)
\times {\bf{E}}_{{\rm{av,T}}}  &=&  i\omega {\bf{B}}_{{\rm{av,T}}}
\l{TAfields}
\end{eqnarray}
where $\varepsilon _h$ is the relative permittivity of the host
medium, and ${\bf{J}}_{d,{\rm{av}}}$ is the averaged microscopic
current. It should be clear that the TA-fields depend exclusively on
$z$ the coordinate.

For the considered wire media, formed exclusively by metallic
inclusions, the averaged microscopic current is given by
\cite{Silv_TAfield}: \e {\bf{J}}_{d,{\rm{av}}} \left( z \right) =
\frac{1}{{A_{cell} }}\int\limits_{\partial A\left( z \right)}
{{\bf{J}}_c \left( {\bf{r}} \right)e^{-i{\bf{k}}_{||} {\bf{.r}}}
\frac{1}{{\left| {{\bf{\hat \nu }} \times {\bf{\hat u}}_z }
\right|}}dl} \l{Jdav}\f where ${\bf{J}}_c  = {\bf{\hat \nu }} \times
{\bf{b}}/\mu _0 $ is the density of electric current on the surface
of a given wire, ${\bf{\hat \nu }}$ is the unit vector normal to the
surface of the wire, ${\partial A\left( z \right)}$ is the contour
determined by the intersection of the surface of the wires in the
unit cell and the pertinent $z=const.$ plane, and $dl$ is the
element of arc. Strictly speaking formula \r{Jdav} is only valid
when there are no wires parallel to the transverse plane (otherwise
the integrand becomes singular), which is the case of interest in
this work. The key result is that, similar to the analysis of Ref.
\cite{ABCtilted}, it is possible to write ${\bf{J}}_{d,{\rm{av}}}$
in terms of the microscopic electric currents induced on the
metallic wires.

Indeed, assuming that the wires are relatively thin, $r_w \ll a$, it
is a good approximation to consider that the density of current over
the $n$-th wire in the unit cell (parallel to the unit vector
${\bf{\hat u}}_n$) is of the form (thin wire approximation): \e
\left. {{\bf{J}}_c } \right|_{\partial D_n } = \frac{{I_n \left( z
\right)}}{{2\pi r_w }}e^{ + i{\bf{k}}_{||} {\bf{.r}}} {\bf{\hat
u}}_n \l{Jc}\f where ${\partial D_n }$ represents the surface of the
considered wire, $I_n$ is the electric current, and $n=1..N$, where
$N$ is the number of components of the wire medium. Notice that the
density of current is modulated by the exponential factor $e^{ +
i{\bf{k}}_{||} {\bf{.r}}}$, because of the assumed Floquet
periodicity of the microscopic fields in the transverse plane.
Substituting Eq. \r{Jc} into Eq. \r{Jdav}, it may be shown that the
averaged microscopic current is given by: \e {\bf{J}}_{d,{\rm{av}}}
\left( z \right) = \frac{1}{{a^2 }}\sum\limits_n {I_n \left( z
\right){\bf{\hat u}}_n } \l{Jdavsimp} \f Thus
${\bf{J}}_{d,{\rm{av}}}$ has indeed a simple relation with the
microscopic electric currents. On the other hand, from Eq.
\r{TAfields} it is straightforward to write the averaged current as
a function of the macroscopic TA-electric field:
\begin{eqnarray}
&&i\omega \mu _0 {\bf{J}}_{d,{\rm{av}}} \left( z \right) = \nonumber
\\
&&\left( {  i{\bf{k}}_{||}  + \frac{d}{{dz}}{\bf{\hat u}}_z }
\right)\left( {  i{\bf{k}}_{||}  + \frac{d}{{dz}}{\bf{\hat u}}_z }
\right){\bf{.E}}_{{\rm{av,T}}}  \nonumber \\
&&+ \left( {k_{||}^2  - \left( {\frac{\omega }{c}} \right)^2
\varepsilon _h  - \frac{{d^2 }}{{dz^2 }}}
\right){\bf{E}}_{{\rm{av,T}}} \l{Jdav_fnE}
\end{eqnarray}
Eqs. \r{Jdavsimp} and \r{Jdav_fnE} establish the connection between
the microscopic currents and the macroscopic electric field and will
be used in what follows to derive additional boundary conditions for
nonconnected wire media.

Let us consider first the case where the material adjacent to the
wire medium is non-conducting (e.g. a dielectric). In such
situation, similar to the case of an array of parallel wires
\cite{MarioABC, ABCtilted}, the microscopic electric current must
vanish at the interface. This means that $I_n=0$, $n=1..N$, at the
interface with the dielectric material, i.e. the currents associated
with different wires in the unit cell must vanish independently.
But, Eq. \r{Jdavsimp} shows that this property implies that the
averaged current verifies: \e{\bf{J}}_{d,{\rm{av}}} {\bf{.\hat u}}_n
= 0,\quad \rm{(diel. \,\, interface)}, \, \emph{n=1..N} \l{ABCair}\f
This equation, along with Eq. \r{Jdav_fnE}, may be regarded as a set
of $N$ independent additional boundary conditions at the interface
with the dielectric material. It should be clear that Eq. \r{ABCair}
may be written exclusively in terms of the macroscopic electric
field and its derivatives, and that the number of additional
boundary conditions is equal to the number of metallic wires in a
unit cell ($N=2$ for the double wire medium and $N=3$ for the triple
wire medium). Obviously, Eq. \r{ABCair} assumes that all the
metallic wires intersect the interface. When some ${\bf{\hat u}}_n$
is parallel to the interface, in general ${\bf{J}}_{d,{\rm{av}}}
{\bf{.\hat u}}_n $ does not vanish at the interface because the path
of the current is not interrupted.

The ABCs \r{ABCair} are a generalization of the results of our
previous works \cite{MarioABC, ABCtilted} for an array of parallel
wires ($N=1$). As demonstrated in \cite{MarioABC, ABCtilted}, when
$N=1$ the ABC and the classical boundary conditions imply the
continuity of the normal component of the electric field multiplied
by the host permittivity at the interface. In particular, when the
host material is air, the ABC, together with the classical boundary
conditions, is equivalent to the continuity of all the Cartesian
components of the electromagnetic field. Interestingly, when $N=2$
or $N=3$ the situation is quite different, specifically, since the
number of ABCs is greater than one, the continuity of the normal
component of the electric field multiplied by the host permittivity
(one single independent equation) cannot be regarded as equivalent
to all the ABCs (two or three independent equations).

Another case of interest occurs when the material adjacent to the
wire medium has very high conductivity (e.g. a metal). It is obvious
that in such configuration the ABCs \r{ABCair} are not valid,
because, assuming that the wires are connected with good Ohmic
contact to the ground plane, the current path is not interrupted at
the interface. In fact, it was demonstrated in Ref. \cite{ABCtilted}
that it is the microscopic electric density of charge $\sigma_c$
that vanishes at the connection point with the metallic surface,
rather than the electric current. Thus, electric charge cannot be
accumulated at the connection points between the wires and the
ground plane. For a wire directed along the $n$-th direction, the
condition $\sigma_c = 0$ is equivalent to $\frac{{dJ_{c,n} }}{{ds}}
= 0$, where $J_{c,n} = \frac{{I_n \left( z \right)}}{{2\pi r_w }}e^{
+ i{\bf{k}}_{||} {\bf{.r}}} $ is the density of current along the
considered wire (see Eq. \r{Jc}), and $s$ is a coordinate measured
along the wire axis. It is clear that for the $n$-th wire
$\frac{d}{{ds}} = {\bf{\hat u}}_n .\nabla$, and hence the condition
$\frac{{dJ_{c,n} }}{{ds}} = 0$ is equivalent to ${\bf{\hat u}}_n
{\bf{.}}\left( {i{\bf{k}}_{||}  + {\bf{\hat u}}_z \frac{d}{{dz}}}
\right)I_n \left( z \right) = 0$. Thus, using Eq. \r{Jdavsimp}, it
is readily found that the averaged current must verify:
\begin{eqnarray}
\left( {i{\bf{k}}_{||}  + {\bf{\hat u}}_z \frac{d}{{dz}}}
\right){\bf{.\hat u}}_n  {\bf{\hat u}}_n {\bf{.J}}_{d,{\rm{av}}} =
0,
\,\, &&\rm{(metallic \,\,\, interface)} \nonumber \\
&&\emph{n=1..N} \l{ABCmetal}
\end{eqnarray}
The above equations define a set of $N$ independent additional
boundary conditions at the interface between a nonconnected wire
medium and a ground plane, and are a generalization of the results
of \cite{ABCtilted}. Again, it should be clear that Eqs.
\r{Jdav_fnE} and \r{ABCmetal} define a functional relation between
the macroscopic electric field and its derivatives at the interface,
and that it is implicit that all the wires intersect the interface.

The application of the proposed ABCs will be illustrated in sections
\ref{SectYOZ} and \ref{SectXOZ}, where, for the case of a double
wire medium, it is shown with full wave simulations that the ABCs
may enable the accurate analytical modeling of wave propagation.

An important point, which is discussed next, is the characterization
the TA- macroscopic fields in the wire medium. The obvious idea is
to write the macroscopic fields in the material as a superposition
of plane waves. The plane waves may be characterized using the
homogenization model proposed in \cite{Silv_MTT_3DWires,
Constantin_WM, MarioEVL}. Specifically, the dielectric function of
the nonconnected wire medium is,
\begin{eqnarray}
\dyad{\varepsilon} \left( {\omega ,{\bf{k}}} \right) &=& \varepsilon
_h \dyad{\bf{I}}  + \sum\limits_{n = 1}^N {\left( {\varepsilon
_{n,n} \left( {\omega ,{\bf{k}} } \right) -
\varepsilon _h } \right){\bf{\hat u}}_n {\bf{\hat u}}_n } \nonumber \\
\varepsilon _{n,n} \left( {\omega ,{\bf{k}}} \right)&=&\varepsilon
_h \left( {1 + \frac{1}{{\frac{1}{{f_V \left( {\varepsilon _{m}
/\varepsilon _h  - 1} \right)}} - \frac{{\left( {\omega /c}
\right)^2 \varepsilon _h  - k_n^2 }}{{\beta _p^2 }}}}} \right)
\nonumber \\ \l{epsWM}
\end{eqnarray}
where $\dyad{\bf{I}}$ is the identity dyadic, ${\bf{\hat u}}_n
{\bf{\hat u}}_n = {\bf{\hat u}}_n  \otimes {\bf{\hat u}}_n$
represents the dyadic (tensor) product of two vectors, $ \beta _p =
\left[ {2\pi /\left( {\ln \left( {a/2\pi r_w } \right) + 0.5275}
\right)} \right]^{1/2} /a$ is the plasma wavenumber, $f_V = \pi
\left( {r_w /a} \right)^2$, $\varepsilon_h$ is the host relative
permittivity, $\varepsilon _m = \varepsilon _m \left( \omega
\right)$ is the metal relative complex permittivity, $c$ is the
speed of light in vacuum, ${\bf{k}} = \left( {k_x ,k_y ,k_z }
\right)$ is the wave vector, and $k_n = {\bf{k}}{\bf{.\hat u}}_n $.

Using the dielectric function \r{epsWM} it is possible to calculate
the $bulk$ macroscopic fields $\left( {{\bf{E}},{\bf{B}}} \right)$.
An important observation is that, for a general metamaterial, the
bulk macroscopic fields may not be coincident with the TA-fields
used in the formulation of the ABCs \r{ABCair} and \r{ABCmetal}.
Indeed, the bulk macroscopic fields, as defined in
\cite{Silv_MTT_3DWires, Silv_TAfield}, are obtained by averaging the
microscopic fields over the unit cell (a volumetric region) of the
periodic material, whereas the TA-fields are obtained by averaging
the microscopic fields over the transverse unit cell (a surface)
\cite{Silv_TAfield}. Fortunately, as demonstrated in Appendix A, for
nonconnected wire media, and supposing that all the wires intersect
the interface (i.e. there are no wires parallel to the interface),
it is possible to identify to a very good approximation the bulk
macroscopic fields with the TA-fields. This means that it is
possible to drop the subscripts ``$\rm{av,T}$'' in Eq. \r{Jdav_fnE},
and characterize the TA-field $\bf{E}_{\rm{av,T}}$ using the
dielectric function of the bulk material. These ideas are further
clarified in sections \ref{SectYOZ} and \ref{SectXOZ}.

\section{Conservation of the power flow \label{SectPowerFlow}}

An important question that may be formulated is if in the lossless
case (i.e. when both $\varepsilon_h$ and $\varepsilon_m$ are real
valued) the proposed ABCs ensure the conservation of the power flow
in the wire medium. Next, we demonstrate that that is indeed the
case. The theory developed below is based on the bulk medium fields,
which as mentioned in the end of the previous section, may be
regarded as equivalent to the transverse averaged fields in the case
of interest.

Let $\left( {{\bf{E}},{\bf{H}}} \right)$ be a solution of Maxwell's
equations in the homogenized bulk wire medium such that the
variation in the $x$ and $y$ coordinates is of the form
$e^{i{\bf{k}}_{||} {\bf{.r}}}$, with ${\bf{k}}_{||}  = \left( {k_x
,k_y ,0} \right)$ a real vector. The bulk magnetic field is by
definition ${\bf{H}}={\bf{B}}/\mu_0$. Thus, $\left(
{{\bf{E}},{\bf{H}}} \right)$ verify:
\begin{eqnarray}
\nabla
\times {\bf{H} } &=& -i\omega {\bf{D}} \nonumber \\
\nabla \times {\bf{E}}  &=&  i\omega \mu_0 {\bf{H}} \l{MaxBulk}
\end{eqnarray}
where ${\bf{D}}= \varepsilon_0 \varepsilon_h {\bf{E}} + {\bf{P}}$ is
the electric displacement vector in the spatially dispersive
material and ${\bf{P}}$ is the nonlocal (generalized) polarization
vector (relative to the host material). Let us define ${\bf{S}}^0$
as follows: \e{\bf{S}}^0 = \frac{1}{2}{\mathop{\rm Re}\nolimits}
\left\{ {{\bf{E}} \times {\bf{H}}^* } \right\} \l{Sz0def}\f It is
well-known that in general ${\bf{S}}^0$ cannot be identified with
the Poynting vector in a spatially dispersive material \cite{Landau,
Agranovich}. In fact, there is no general formula for the Poynting
vector in a spatially dispersive material. The only case for which
the Poynting vector is actually known is when the material is
lossless, and the electromagnetic field is associated with a
propagating plane wave \cite{Agranovich}. In what follows, we will
demonstrate that for lossless nonconnected wire media, it is
possible to define unambiguously the $z$-component of the Poynting
vector, even if the electromagnetic field has an arbitrary variation
with $z$ (not necessarily a propagating plane-wave).

To begin with, we note that from Eq. \r{MaxBulk} it is immediate
that: \e \nabla {\bf{.S}}^0  = \frac{1}{2}{\mathop{\rm Re}\nolimits}
\left\{ { - i\omega {\bf{E}}{\bf{.D}}^* }
 \right\}  = \frac{1}{2}{\mathop{\rm Re}\nolimits} \left\{ { - i\omega {\bf{E}}{\bf{.P}}^* } \right\} \l{divS0}\f
where the second identity is a consequence of the fact that the host
permittivity $\varepsilon_h$ is a real number in the absence of
loss. The next step is to relate $\bf{P}$ and $\bf{E}$. In the
spectral (Fourier) domain we clearly have that: \e {\bf{\tilde
P}}\left( {\bf{k}} \right) = \varepsilon _0 \left( {\dyad{
\varepsilon } \left( {\omega ,{\bf{k}}} \right) - \varepsilon _h
\dyad{{\bf{I}}} } \right){\bf{.\tilde E}}\left( {\bf{k}} \right) \f
where $\dyad{\varepsilon }$ is the relative permittivity of the wire
medium, the symbol ``$\sim$'' denotes the spatial Fourier transform,
and here $\bf{k}$ represents the Fourier coordinates. In particular,
the projections of $\bf{\tilde E}$ and $\bf{\tilde P}$ onto the
principal direction ${\bf{\hat u}}_n$ of the dielectric function
verify $\tilde P_n  = \varepsilon _0 \left( {\varepsilon _{n,n}
\left( {\omega ,k_n } \right) - \varepsilon _h } \right)\tilde E_n$
with $k_n  = {\bf{k}}{\bf{.\hat u}}_n$ ($n=1..N$). Thus, using Eq.
\r{epsWM} it follows that: \e \tilde E_n  = \frac{1}{{\varepsilon _0
\varepsilon _h \beta _p^2 }}\left( {k_n^2  - \frac{{\omega ^2
}}{{c^2 }}\varepsilon _h  + \frac{{\beta _p^2 }}{{f_V \left(
{\varepsilon _m /\varepsilon _h  - 1} \right)}}} \right)\tilde P_n
\f Hence, calculating the inverse Fourier transform, we obtain the
following relation in the space domain: \e E_n  =
\frac{1}{{\varepsilon _0 \varepsilon _h \beta _p^2 }}\left[ {\left(
{ - i{\bf{\hat u}}_n {\bf{.}}\nabla } \right)^2  - \frac{{\omega ^2
}}{{c^2 }}\varepsilon _h  + \frac{{\beta _p^2 }}{{f_V \left(
{\varepsilon _m /\varepsilon _h  - 1} \right)}}
 } \right]P_n \f
But since, it is assumed that the variation of the electromagnetic
field in the $x$ and $y$ coordinates is of the form
$e^{i{\bf{k}}_{||} {\bf{.r}}}$, we have that $
 - i{\bf{\hat u}}_n {\bf{.}}\nabla  = {\bf{\hat u}}_n {\bf{.k}}_{||}  - i{\bf{\hat u}}_n {\bf{.\hat u}}_z
 \frac{d}{{dz}}$. Thus, after some simplifications, it is found that:
\begin{eqnarray}
{\mathop{\rm Re}\nolimits} \left\{ { - iP_n^* E_n } \right\} &=&
\frac{1}{{\varepsilon _0 \varepsilon _h \beta _p^2 }}{\mathop{\rm
Re}\nolimits} \left\{ {i\left( {{\bf{\hat u}}_n {\bf{.\hat u}}_z }
\right)^2 P_n^* \frac{{d^2 P_n }}{{dz^2 }}} \right. \nonumber \\
&&{\left. { - 2\left( {{\bf{k}}_{||} {\bf{.\hat u}}_n }
\right)\left( {{\bf{\hat u}}_n {\bf{.\hat u}}_z } \right)P_n^*
\frac{{dP_n }}{{dz}}} \right\}}
\end{eqnarray}
Next we note that ${\mathop{\rm Re}\nolimits} \left\{ {i{\kern 1pt}
P_n^* \frac{{d^2 P_n }}{{dz^2 }}} \right\} = {\mathop{\rm
Re}\nolimits} \left\{ {\frac{d}{{dz}}\left( {i{\kern 1pt} P_n^*
\frac{{dP_n }}{{dz}}} \right)} \right\}$ and that ${\mathop{\rm
Re}\nolimits} \left\{ {P_n^* \frac{{dP_n }}{{dz}}} \right\} =
\frac{d}{{dz}}\left( {\frac{{\left| {P_n } \right|^2 }}{2}} \right)$
to finally write:
\begin{eqnarray}
{\mathop{\rm Re}\nolimits} \left\{ { - iP_n^* E_n } \right\} &=&
\frac{1}{{\varepsilon _0 \varepsilon _h \beta _p^2
}}\frac{d}{{dz}}{\mathop{\rm Re}\nolimits} \left\{ {\left(
{{\bf{\hat u}}_n {\bf{.\hat u}}_z } \right)^2 i{\kern 1pt} P_n^*
\frac{{dP_n }}{{dz}}} \right.
 \nonumber \\
&&\left. { - \left( {{\bf{k}}_{||} {\bf{.\hat u}}_n } \right)\left(
{{\bf{\hat u}}_n {\bf{.\hat u}}_z } \right)\left| {P_n } \right|^2 }
\right\} \l{ReEnPn}
\end{eqnarray}

We are now in a position to calculate the divergence of
${\bf{S}}^0$, given by Eq. \r{divS0}. On one hand, we note that due
to the assumed dependence of $\left( {{\bf{E}},{\bf{H}}} \right)$ on
the $x$ and $y$ coordinates, it is clear that ${\bf{S}}^0$ depends
exclusively on $z$, and thus $\nabla {\bf{.S}}^0 = \frac{{dS_z^0
}}{{dz}}$ where $S_z^0$ is the $z$ component of ${\bf{S}}^0$. On the
other hand, it is obvious that ${\bf{P}}^* {\bf{.E}} =
\sum\limits_{n = 1}^N {P_n^* E_n }$ because the only non-vanishing
components of $\bf{P}$ are precisely the $P_n$'s (remember also that
the unit vectors ${\bf{\hat u}}_n$ are mutually orthogonal).
Therefore Eqs. \r{divS0} and \r{ReEnPn} show that, \e \frac{{dS_z
}}{{dz}} = 0 \l{conservSz}\f where $S_z$ is given by:
\begin{eqnarray}
&&S_z = S_z^0  - \frac{\omega }{2}\frac{1}{{\varepsilon _0
\varepsilon _h \beta _p^2 }} \times \nonumber \\
&&{\mathop{\rm Re}\nolimits} \left\{ {i\sum\limits_{n = 1}^N
{{\bf{\hat u}}_n {\bf{.\hat u}}_z } P_n^* \left( {i{\bf{k}}_{||}  +
\frac{d}{{dz}}{\bf{\hat u}}_z } \right){\bf{.\hat u}}_n P_n }
\right\} \l{Sz}
\end{eqnarray}
As discussed below $S_z$ is the $z$ component of the Poynting vector
in the spatially dispersive material. Before that discussion, let us
show that the conservation law \r{conservSz} implies that the ABCs
introduced in this work guarantee the conservation of power flow at
an interface with a dielectric or a metallic material.

Indeed, it is clear from Eq. \r{Sz} that at the points $z$ such that
for every $n=1..N$, \e {\rm{either}}\; P_n=0 \quad{\rm{or}}\; \left(
{i{\bf{k}}_{||}  + \frac{d}{{dz}}{\bf{\hat u}}_z } \right){\bf{.\hat
u}}_n P_n =0 \l{ABCPn}\f we have that $S_z = S_z^0$. Comparing Eqs.
\r{TAfields} with Eqs. \r{MaxBulk}, and using the fact that in the
nonconnected wire medium the bulk medium fields can be identified
with the TA-fields (see Appendix A), it is evident that the
polarization vector verifies, \e {\bf{P}} = \frac{1}{{ - i\omega
}}{\bf{J}}_{d,{\rm{av}}} e^{i{\bf{k}}_{||} {\bf{.r}}} \f where
${\bf{J}}_{d,{\rm{av}}}$ is the averaged microscopic current. But
then it follows that the first condition in Eq. \r{ABCPn} is
equivalent to the ABC \r{ABCair}, whereas the second condition is
equivalent to the ABC \r{ABCmetal}. Hence, we conclude that at the
points $z$ of the material where either the ABCs \r{ABCair}
(associated with dielectric interfaces) or the ABCs \r{ABCmetal}
(associated with metallic interfaces) are enforced, we have that
$S_z = S_z^0$.

Let's now consider a truncated wire medium (wire medium slab) with
interfaces at $z=z_i$ and $z=z_f$. From Eq.\r{conservSz}, in the
lossless case $S_z$ is constant inside the wire medium, and
consequently when the ABCs are enforced, $S_z^0$ has the same value
at the two interfaces: $S_z^0(z_i)=S_z^0(z_f)$ (calculated from the
wire medium side). But due to the classical boundary conditions
(continuity of the tangential components of $\bf{E}$ and $\bf{H}$ at
the interfaces), $S_z^0$ may be evaluated either at the wire medium
side of the interface or at the exterior side, being the result the
same. Thus, we have demonstrated that the proposed ABCs together
with the classical boundary conditions imply that
$S_z^0(z_i)=S_z^0(z_f)$, being $S_z^0$ evaluated at the exterior
side of the interface, where it is obviously the $z$ component of
the Poynting vector. This shows that in the lossless case the ABCs
\r{ABCair} and \r{ABCmetal} (or more generally the condition
\r{ABCPn}) guarantee, in fact, the conservation of the power flow
through a wire medium slab, as we wanted to prove. This result is
valid for simple ($N=1$), double ($N=2$) or triple ($N=3$) wire
media.

To show that $S_z$ can be identified with the $z$ component of the
Poynting vector in the spatially dispersive material, we consider
the typical case where the electromagnetic field is a superposition
of plane waves: \e {\bf{E}}\left( {\bf{r}} \right) = \sum\limits_l
{{\bf{E}}_l e^{ + i{\kern 1pt} {\bf{k}}_l {\bf{.r}}} }\l{PWs}\f
where ${\bf{E}}_l$ is a constant vector (determines the polarization
of the plane wave), and ${\bf{k}}_l = {\bf{k}}_{||}  + k_z^{\left( l
\right)} {\bf{\hat u}}_z$ is the wave vector associated with the
plane wave. The $z$-component $k_z^{\left( l \right)}$ may be either
real (positive or negative) or complex valued (as discussed in
section \ref{sectNum1}, the wire medium may support evanescent modes
even in the absence of loss). It is supposed that each individual
plane wave verifies Eqs. \r{MaxBulk}. In Appendix B, we demonstrate
that for such superposition of plane waves, $S_z$, defined by
\r{Sz}, may be rewritten as:
\begin{eqnarray}
S_z  = \sum\limits_{\scriptstyle l,m \hfill \atop
  \scriptstyle {\bf{k}}_l  = {\bf{k}}_m^*  \hfill} {{\mathop{\rm Re}\nolimits} \left\{ {\left( {\frac{1}{2}{\bf{E}}_l  \times {\bf{H}}_m^* } \right){\bf{.\hat u}}_z } \right.}
 \nonumber \\
\left. { - \frac{{\omega \varepsilon _0 }}{4}{\bf{E}}_m^*
{\bf{.}}\frac{{\partial \dyad{ \varepsilon } \left( {\omega
,{\bf{k}}_l } \right)}}{{\partial k_z }}{\bf{.E}}_l } \right\}
\l{SzPW}
\end{eqnarray}
where the summation is restricted to the indices $l,m$ such that
${\bf{k}}_l  = {\bf{k}}_m^*$, and ${\bf{H}}_l$ is the magnetic field
associated with ${\bf{E}}_l$. But for a single plane wave with real
wave vector the above formula reduces to the well-known expression
of the Poynting vector in a general spatially dispersive material
\cite{Landau, Agranovich}. This demonstrates that $S_z$ may indeed
be identified with the $z$-component of the Poynting vector. It
should however be stressed that the above formula is more general
than the results reported in \cite{Landau, Agranovich}, which apply
exclusively to a single plane wave with real wave vector. The result
\r{SzPW} generalizes the classical result to the case of a
superposition of plane waves (possibly associated with complex wave
vectors) in a lossless nonconnected wire medium.

\section{Crossed Wire Mesh \label{sectNum1}}

\begin{figure}[th] \centering
\epsfig{file=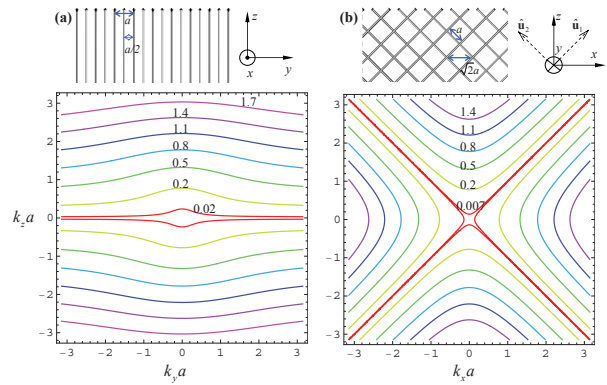, width=8.0cm} \caption{(Color
online) Isofrequency contours of the fundamental plane wave mode for
(a) propagation in the $yoz$ plane with electric field along the
$x$-direction; (b) propagation in the $xoz$ plane with electric
field in the same plane. The radius of the wires is $r_w=0.05a$ and
the metal is assumed perfectly conducting. The text insets indicate
the value of the normalized frequency $\omega a/c$.} \label{Isofreq}
\end{figure}

In the rest of the paper we apply the developed theory to the
particular case of a crossed wire mesh formed by a double array of
metallic wires \cite{MarioEVL, IgorPRE, Constantin_WM}. It is
assumed that the wires are parallel to the $xoz$-plane, and that
${\bf{\hat u}}_1 = {{\left( {1,0,1} \right)}/{\sqrt 2 }}$ and
${\bf{\hat u}}_2 = {{\left( {-1,0,1} \right)}/{\sqrt 2 }}$. Thus,
the angle of the wire axes with respect to the interface normal is
$\pm45^o$. A cut of the considered structure in the planes $yoz$ and
$xoz$ is represented in Fig. \ref{Isofreq}. We will analyze the
scattering and guiding of electromagnetic waves by the wire medium
slab in the cases where the wave vector is confined to one of these
two planes. Moreover, in case of propagation in the $yoz$ plane with
will assume that the electric field is polarized along the
$x$-direction, whereas in case of propagation in the $xoz$-plane we
will restrict our attention to the situation where the magnetic
field is along the $y$-direction (thus, in both cases, the electric
field is in the $xoz$ plane).

Before addressing the problem of propagation in a finite structure
(wire medium slab), next we will briefly review some key properties
of the electromagnetic modes in the unbounded crossed wire mesh.
Unlike a conventional local plasma, the metamaterial may support
propagating modes for arbitrarily low frequencies \cite{MarioEVL,
Silv_MTT_3DWires, Constantin_WM, IgorPRE, Shin}. Specifically, in
the lossless case (e.g. for perfect electrical conductors (PEC),
i.e. $\varepsilon_m=-\infty$), and for frequencies much lower than
the effective plasma frequency, $\omega /c \ll \beta _p /\sqrt
{\varepsilon _h }$, the wire medium supports a propagating plane
wave with electric field in the $xoz$ plane.

Following Refs. \cite{MarioEVL, XwiresLens}, the plane waves
supported by the material in case of propagation in the $yoz$ plane
($k_x=0$) with electric field along the $x$-direction are
characterized by the dispersion characteristic:
\begin{eqnarray}
&&{\rm{\quad}}\varepsilon \left( {\omega ,k_z } \right)\left(
{\omega /c}
\right)^2  = k_y^2  + k_z^2, {\rm{\quad with}} \nonumber \\
&&\varepsilon\left( {\omega ,k_z} \right)=\varepsilon _h \left( {1 +
\frac{1}{{\frac{1}{{f_V \left( {\varepsilon _{m} /\varepsilon _h  -
1} \right)}} - \frac{{\left( {\omega /c} \right)^2 \varepsilon _h  -
k_z^2/2 }}{{\beta _p^2 }}}}} \right) \nonumber \\ \l{dispyoz}
\end{eqnarray}
The isofrequency contours of the propagating mode are depicted in
Fig.\ref{Isofreq}a, for PEC wires with normalized radius
$r_w/a=0.05$. Clearly, the material is strongly anisotropic, and the
isofrequency contours remind ellipses with a large axial ratio. It
is seen that the contours are nearly perpendicular to the
$z$-direction, and consequently the energy tends to flow along $z$.
This suggests that adjacent wire planes tend to guide the wave along
$z$, and obstruct the propagation along $y$. It was shown in Ref.
\cite{MarioEVL} that the effective index of refraction for
propagation along $z$ is (assuming $\varepsilon_h=1$ and PEC wires):
\e n_{ef} \equiv \left. {\frac{{k_z c}}{\omega }} \right|_{k_y = 0}
= \sqrt {\frac{3}{2} + \frac{1}{2}\left( {1 + 8\left( {\frac{{\beta
_p c}}{\omega }} \right)^2 } \right)^{1/2} } \l{nref}\f Since $\beta
_p \sim 1/a$, one interesting feature of this structure is that for
a fixed frequency the index of refraction may be greatly enhanced by
reducing the spacing between the wires (maintaining the metal volume
fraction) \cite{MarioEVL}. Other mechanisms to design materials with
a high-index were proposed in \cite{Shen2005, Shin2009}, but the
required microstructures are arguably much more complicated to
fabricate than a crossed wire mesh.

As could be expected from the geometry of the material, the
propagation properties in the $xoz$ plane ($k_y=0$) are very
different. The dispersion characteristic is now \cite{XWiresNegRef}:
\e \frac{{k_1^2 }}{{k^2  - \left( {\omega /c} \right)^2 \varepsilon
_{11} }} + \frac{{k_2^2 }}{{k^2  - \left( {\omega /c} \right)^2
\varepsilon _{22} }} = 1 \l{dispEqXOZ}\f where $\varepsilon _{n,n}$
($n=1,2$) is given by Eq. \r{epsWM}, $k^2=k_x^2+k_z^2$, and $k_n  =
{\bf{k}}{\bf{.\hat u}}_n$, with ${\bf{k}} = \left( {k_x ,0,k_z }
\right)$. The corresponding isofrequency contours are depicted in
Fig.\ref{Isofreq}b, and remind two perpendicular hyperbolas. The
electric field associated with each plane wave is nearly (but not
exactly) tangent to the isofrequency contours \cite{XWiresNegRef}.
The shape of the isofrequency contours suggests that propagation is
favored along the coordinates axes, i.e. when the wave vector makes
an angle of $\pm45^o$ with the wire axes. On the other hand, when
the wave vector is normal to one of wire axes the structure has a
directional bandgap. The hyperbolic shaped isofrequency contours may
enable the emergence of negative refraction \cite{XWiresNegRef,
Constantin_WM, IgorPRE}, as will be further discussed in section
\ref{SectXOZ}.

\section{Propagation in the \emph{yoz} plane \label{SectYOZ}}

In what follows, we apply the ABCs derived in section \ref{ABC} to
characterize the scattering of plane waves by a double wire medium
slab, assuming propagation in the $yoz$ plane (Fig.\ref{Isofreq}a).
As in section \ref{ABC}, it is assumed that the normal to the
interfaces is along the $z$-direction.

To begin with, we calculate the averaged microscopic current in the
wire medium slab. It is assumed that the excitation is such that the
electric field is along the $x$-direction. Clearly, for propagation
in the $yoz$-plane the transverse wave vector is such that
${\bf{k}}_{||}  = k_y {\bf{\hat u}}_y$, where $k_y$ is determined by
the excitation (for an incoming plane wave, $k_y = \omega/c
\sin\theta_i$ where $\theta_i$ is the angle of incidence). Thus,
from Eq. \r{Jdav_fnE}, it is found that the averaged current
verifies, \e i\omega \mu _0 {\bf{J}}_{d,{\rm{av}}} \left( z \right)
= \left( {k_y^2  - \varepsilon _h \frac{{\omega ^2 }}{{c^2 }} -
\frac{{d^2 }}{{dz^2 }}} \right)E_x {\bf{\hat u}}_x \l{Jdsimp}\f
where $E_x$ is the electric field in the wire medium slab. The ABCs
at a dielectric interface can now be easily obtained using Eq.
\r{ABCair}. As discussed in section \ref{ABC}, in the double wire
medium case, one needs to impose two distinct ABCs at the interface
($N=2$). However, in the present case the two ABCs \r{ABCair} impose
the same condition on the macroscopic electric field: \e {\frac{{d^2
E_x }}{{dz^2 }} + \left(\varepsilon _h \frac{{\omega ^2 }}{{c^2
}}-k_y^2 \right) E_x } = 0 \quad \rm{(diel. \,\,
interface)}\l{ABCdielYOZ}\f This degeneracy occurs due to the
exceptionally high-symmetry of the system. In the general case, the
ABCs are not redundant and yield, in fact, independent equations, as
will be shown in Sect. \ref{SectXOZ}. Similarly, at a metallic
interface, the two ABCs \r{ABCmetal} also yield a single equation:
\e\frac{{d^3 E_x }}{{dz^3 }} + \left( {\varepsilon _h \frac{{\omega
^2 }}{{c^2 }} - k_y^2 } \right)\frac{{dE_x }}{{dz}} = 0 \quad
\rm{(metallic \,\, interface)} \l{ABCmetYOZ}\f The equations
\r{ABCdielYOZ} and \r{ABCmetYOZ} correspond exactly to the
additional boundary conditions that were enunciated in our previous
works \cite{MarioEVL, XwiresLens} (with no proof). These ABCs can be
readily applied to solve a scattering or wave guiding problem when
the wave vector is confined to the $yoz$ plane. Since the use of the
ABCs for this specific configuration was already addressed in Refs.
\cite{MarioEVL, XwiresLens}, our discussion here will be somehow
abbreviated.

\subsection{Slab standing in free-space \label{SubSectYOZFS}}

\begin{figure}[th] \centering
\epsfig{file=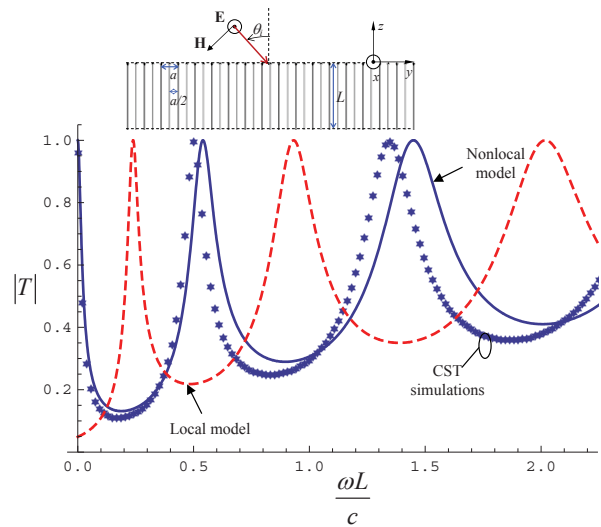, width=8.0cm} \caption{(Color
online) Magnitude of the transmission coefficient as a function of
frequency for a metamaterial slab with a lattice constant $a=L/15$
and $r_w=0.05a$ and a fixed thickness $L$. The solid line represents
the spatially dispersive (nonlocal) model, the dashed line
represents the local model, and the discrete symbols were calculated
with CST Microwave Studio. The inset shows the geometry of the
problem.} \label{EVLlocal}
\end{figure}

First we consider the case where the wire medium slab stands in
free-space, and is illuminated by a plane wave characterized by the
transverse wave number $k_y = \omega/c \sin\theta_i$ (see the inset
of Fig. \ref{EVLlocal}). As mentioned in Sect. \ref{ABC}, the key
assumption is that the electromagnetic field inside the wire medium
slab can be written as superposition of plane waves characterized by
the dielectric function of the bulk (unbounded) material. Because of
translational invariance $k_y$ must be conserved, and thus the
pertinent plane waves can be found from the solution of the
dispersion characteristic \r{dispyoz} with respect to $k_z$, with
$\omega$ and $k_y$ determined by the excitation. As discussed in
Refs. \cite{MarioEVL, XwiresLens}, the solution of Eq. \r{dispyoz}
yields exactly two different solutions for $k_z^{2}$. It should be
clear that for a regular local dielectric slab there is only one
single solution. This clarifies why for propagation in the $yoz$
plane, a single ABC is sufficient to characterize interface effects
(despite that the wire medium has two components), i.e. since there
is only one additional wave only one ABC is required. Thus, the
electric field inside the the metamaterial slab may be written as
(the $y$-dependence of the field is suppressed):
\begin{eqnarray}
E_x &=&  {A^{+}_1  e^{ +i {\kern 1pt} k_{z,(1)} z}  + A^{-}_1  e^{
-i{\kern 1pt} k_{z,(1)} z} +
} \nonumber \\
&& {A^{+}_2  e^{ +i {\kern 1pt} k_{z,(2)} z}  + A^{-}_2 e^{ -i{\kern
1pt} k_{z,(2)} z} } \l{ExfieldYOZ}
\end{eqnarray}
where $k_{z,(i)}= k_{z,(i)} \left( {\omega ,k_y } \right) $ are the
solutions of the dispersion equation \r{dispyoz}. As discussed in
Refs. \cite{MarioEVL, XwiresLens}, for long wavelengths and in the
absence of loss, there is only one propagating mode in the wire
medium, i.e. one of the propagation constants, let's say
$k_{z,(1)}$, is real valued and is associated with the isofrequency
contours depicted in Fig. \ref{Isofreq}a, while the other
propagation constant, $k_{z,(2)}$, is pure imaginary.

The electric field in the air regions (below and above the
metamaterial slab) can be written as well as a superposition of
(free-space) plane waves. By enforcing the continuity of the
tangential components of the electric and magnetic field, and the
ABC \r{ABCdielYOZ} it is possible to determine the reflection and
transmission coefficients for plane wave incidence. For further
details the reader is referred to Ref. \cite{XwiresLens} [in
particular, formula (5) of Ref. \cite{XwiresLens} gives a closed
form expression for the transmission coefficient $T$].

In the first example, we consider a metamaterial slab formed by PEC
wires, and such that the spacing between the wires is $a=L/15$,
where $L$ is the thickness of the slab. The host material is air.
The amplitude of the transmission coefficient for plane wave
incidence along $\theta_i = 0.1^o$ is plotted in Fig. \ref{EVLlocal}
as a function of the normalized frequency. It is seen that the
results obtained with the nonlocal homogenization model and the
proposed ABCs (solid line) compare well with the results obtained
with the full wave electromagnetic simulator CST Microwave Studio
(discrete symbols) \cite{CST2008}.

As mentioned before, in the absence of loss, the propagation
constant $k_{z,(1)}$ may be assumed real valued (associated with a
propagating mode in the wire medium), whereas the propagation
constant of $k_{z,(2)}$ is pure imaginary (associated with an
evanescent mode). Thus, it might be thought that this later
evanescent mode would play a minor role in the response of the
metamaterial slab. In order to test this hypothesis, we have
calculated the transmission coefficient obtained by setting
$A^{\pm}_2 = 0$ in Eq. \r{ExfieldYOZ}. This condition removes the
extra degree of freedom of the problem, and thus to calculate the
transmission coefficient it is sufficient to impose the classical
boundary conditions at the interfaces, i.e. if one neglects the
effect of the evanescent wave it is not necessary to consider an ABC
at the interfaces. We refer to the results obtained with this
approximation, as the ``local model'' results. A bit surprisingly,
as reported in Fig. \ref{EVLlocal}, the results obtained with the
local model (dashed line) are qualitatively very different from the
results obtained with the nonlocal model. In particular, the
frequencies for which the local model predicts the maxima of $T$
correspond to the minima of the nonlocal model, and vice-versa. This
demonstrates that the effect of the evanescent mode and the ABCs
cannot be neglected, and are essential to describe accurately the
response of the material. It should be noted that the evanescent
mode, even though strongly attenuated in the interior of the
metamaterial slab, may be excited in the close vicinity of the
interfaces in order to ensure that the microscopic current in the
metallic wires vanishes at the interfaces. Thus, even though the
wave propagation in the interior of the slab is dominated by the
propagating mode, the ``effective wave impedance'' at the interfaces
is determined by both the propagating and the evanescent modes. This
explains the disagreement between the local and nonlocal models
reported in Fig. \ref{EVLlocal}.

One of the interesting properties of the crossed wire mesh is that
it may interact very strongly with the incoming wave, even when the
the length of the metallic wires is electrically short, effectively
behaving as a material with very large dielectric constant
\cite{XwiresLens}. For instance, in the example depicted in Fig.
\ref{EVLlocal} the first dip of the transmission coefficient occurs
for $\omega L/c \approx 0.2$, which corresponds to the metallic
wires with length $L_w = \sqrt2 L = 0.04 \lambda_0$. This value is
one order of magnitude smaller than the traditional $\lambda_0/2$
resonance of a metallic wire. Moreover, as discussed in Ref.
\cite{XwiresLens}, the resonance length of the wires may be made
arbitrarily small by increasing the density of wires (number of
wires per unit of volume), keeping the metal volume fraction
unchanged. Indeed, unlike conventional metamaterial designs, the
effective index of refraction of the material does not saturate when
the inclusions are scaled and the lattice constant $a$ is made
smaller.
\begin{figure}[th] \centering
\epsfig{file=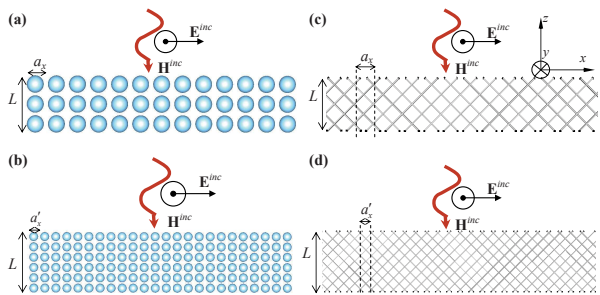, width=8.0cm} \caption{(Color online)
Panel (a): generic metamaterial slab formed by inclusions whose size
is smaller than the lattice constant. Panel (c): crossed wire mesh
slab formed by inclusions that are spanned over many unit cells. As
discussed in the main text, unlike a standard metamaterial, the
response of the crossed wire mesh does not saturate when the density
of inclusions (number of inclusions per unit of volume) is
increased, keeping the volume fraction constant (panels (b) and
(d)).} \label{Compar}
\end{figure}

To further clarify these ideas we consider the scenarios depicted in
Fig. \ref{Compar}. In panels (a) and (b) we represent a conventional
metamaterial slab (e.g. an array of metal or dielectric spheres)
with fixed thickness $L$, but different lattice constants $a_x$. The
volume fraction of the inclusions is assumed to be the same in the
two cases. Each inclusion is entirely contained in a basic cell. The
incoming wave illuminates the slab along the normal direction. From
the classical Clausius-Mossotti formula \cite{Jackson}, it should be
clear that in the quasi-static limit the value of the index of
refraction of the bulk metamaterial will eventually saturate and
become independent of the lattice constant $a_x$, when $a_x$ is made
sufficiently small (keeping the volume fraction of the inclusions
unchanged). In fact, for a fixed frequency, and for $a_x $
increasingly small, the index of refraction will obviously converge
to the static case value. In particular, it is expected that in the
quasi-static limit the transmission coefficient in configurations
(a) and (b) becomes independent of $a_x$, provided $a_x$ is
sufficiently small.

The situation for a crossed wire mesh is completely different. As
discussed in Sect. \ref{sectNum1} - see Eq. \r{nref} - for a fixed
frequency and a fixed metal volume fraction, the index of refraction
of the bulk wire mesh can be made arbitrarily large by reducing the
lattice constant $a_x$. Thus, despite the fact that the volume
fraction of the metal is kept constant, the response of the
structures depicted in Fig. \ref{Compar}c and Fig. \ref{Compar}d may
be radically different. In particular, unlike in a conventional
metamaterial, the response does not saturate when $a_x$ is made
increasingly small. This important qualitative difference between
the wire mesh and a conventional metamaterial stems from the fact
that in a wire mesh each inclusion is spanned over many unit cells,
i.e. each individual wire crosses several unit cells (see Figs.
\ref{Compar}c and \ref{Compar}d). Obviously, such property precludes
the application of the Clausius-Mossotti formula, and clarifies the
strong coupling between the different inclusions and the anomalous
electric response of the wire mesh.

To illustrate how the response of the wire mesh is modified when the
density of wires is increased, we have calculated the dispersion
characteristic of the guided modes supported by the metamaterial
slab (i.e. the waves that may propagate along the $y$-direction
attached to the interfaces, even in the absence of an external
source). The dispersion characteristic of the guided modes, $k_y  =
k_y \left( \omega \right)$, may be obtained from the poles of the
transmission coefficient for plane wave incidence \cite{XwiresLens}.
\begin{figure}[th] \centering
\epsfig{file=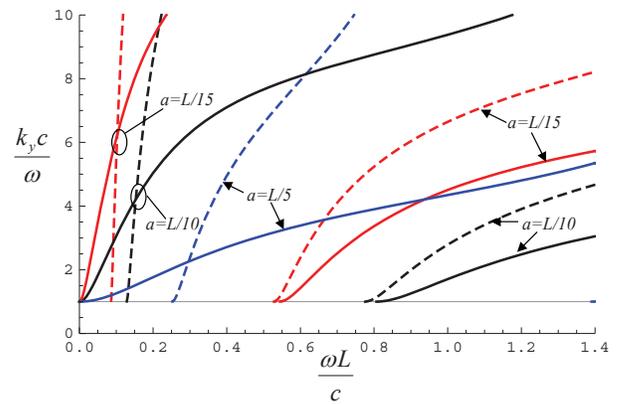, width=8.0cm} \caption{(Color online)
Dispersion characteristic of surface waves for a metamaterial slab
with thickness $L$ and different values of the lattice constant $a$.
The radius of the wires is $r_w=0.05a$. Solid lines: metamaterial
slab stands in free-space. Dashed lines: grounded metamaterial
slab.} \label{sws}
\end{figure}

In Fig. \ref{sws} we depict (solid lines) the calculated dispersion
characteristics for a slab with thickness $L$ and different values
of the lattice constant $a$. The metal volume fraction is constant
in all the examples ($r_w=0.05a$). The results were obtained using
the homogenization model. Similar to a conventional dielectric
substrate, the crossed wire mesh supports a guided mode with no
cut-off frequency, i.e. for arbitrarily long wavelengths. Consistent
with the previous discussions, it can be seen that for a fixed
frequency the effective index of refraction of the guided mode, $k_y
c/\omega$, increases significantly when the lattice constant $a$ is
reduced from $a=L/5$ to $a=L/15$, i.e. when the density of the wires
is increased. This simple example illustrates how unlike a
conventional metamaterial slab, the response of the crossed wire
mesh does not saturate as the lattice constant is made increasingly
smaller.

Even though, for simplicity, the examples considered here assume PEC
wires, the proposed homogenization model can be applied as well when
either the effect of loss or the plasmonic response of the metal are
taken into account. For further details, the reader is referred to
Ref. \cite{XwiresLens}, where the model has been validated against
full wave simulations in such scenarios.

\subsection{Grounded Slab \label{SubSectYOZGround}}

\begin{figure}[th] \centering
\epsfig{file=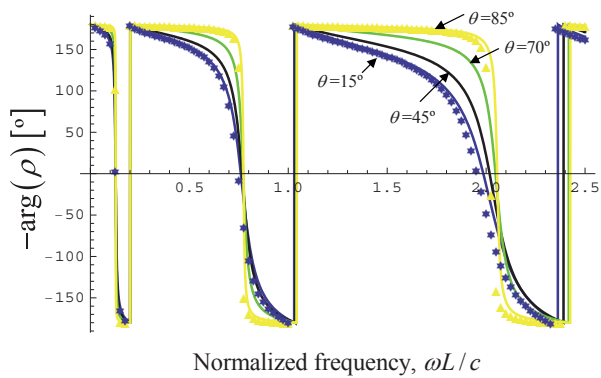, width=8.0cm} \caption{(Color online)
Phase of the reflection coefficient as a function of the normalized
frequency for different angles of incidence. The lattice constant is
$a = L/10$ and the radius of the wires is $r_w=0.05a$, where $L$ is
the thickness of the grounded slab. The host material is air. Solid
lines: Homogenization model. Star and triangle shaped symbols: full
wave simulations for $\theta=15^o$ and $\theta=85^o$, respectively.}
\label{phRevl}
\end{figure}

Next, we consider the case in which the bottom face of the
metamaterial slab is covered with a metallic plane \cite{MarioEVL}.
It is supposed that the metallic wires are attached to the ground
plane with good Ohmic contact. In this configuration, assuming plane
wave incidence with electric field polarized along the
$x$-direction, the electric field inside the metamaterial slab can
still be written as in Eq. \r{ExfieldYOZ}. The boundary conditions
at the air interface are the same as in Sect. \ref{SubSectYOZFS}. On
the other hand, at the metallic interface $E_x$ must vanish, and the
additional boundary condition \r{ABCmetYOZ} must be enforced.
Proceeding in this manner (see Ref. \cite{MarioEVL}), it may be
proven that the reflection coefficient referred to the air interface
is:
\begin{widetext}
\begin{eqnarray}
&&\rho =  - 1 + \frac{{2\gamma _0 \left( {k_{z,\left( 2 \right)}^2 -
k_{z,\left( 1 \right)}^2 } \right)}}{{D\left( {\omega ,k_y }
\right)}}\left[ {k_{z,\left( 2 \right)} \left( {\gamma _h^2  +
k_{z,\left( 2 \right)}^2 } \right)\tan \left( {k_{z,\left( 1
\right)} L} \right) - k_{z,\left( 1 \right)} \left( {\gamma _h^2  +
k_{z,\left( 1 \right)}^2 } \right)\tan \left( {k_{z,\left( 2
\right)} L} \right)} \right] \l{rohYOZ}\\
 &&D\left( {\omega ,k_y } \right) =
k_{z,\left( 1 \right)} k_{z,\left( 2 \right)} \left[ {2\gamma _h^2
\left( {\gamma _h^2  + k_{z,\left( 1 \right)}^2  + k_{z,\left( 2
\right)}^2 } \right) + k_{z,\left( 1 \right)}^4  + k_{z,\left( 1
\right)}^4 } \right] +
 \nonumber \\
 && \left( {\gamma _h^2  + k_{z,\left( 1 \right)}^2 } \right)\left( {\gamma _h^2  + k_{z,\left( 2 \right)}^2 } \right)\left[ {\left( {k_{z,\left( 1 \right)}^2  + k_{z,\left( 2 \right)}^2 } \right)\tan \left( {k_{z,\left( 1 \right)} L} \right)\tan \left( {k_{z,\left( 2 \right)} L} \right) - 2k_{z,\left( 1 \right)} k_{z,\left( 2 \right)} \sec \left( {k_{z,\left( 1 \right)} L} \right)\sec \left( {k_{z,\left( 2 \right)} L} \right)}
 \right] + \nonumber \\
&& \gamma _0 \left( {k_{z,\left( 2 \right)}^2  - k_{z,\left( 1
\right)}^2 } \right)\left[ {k_{z,\left( 2 \right)} \left( {\gamma
_h^2  + k_{z,\left( 2 \right)}^2 } \right)\tan \left( {k_{z,\left( 1
\right)} L} \right) - k_{z,\left( 1 \right)} \left( {\gamma _h^2  +
k_{z,\left( 1 \right)}^2 } \right)\tan \left( {k_{z,\left( 2
\right)} L} \right)} \right] \l{dispPECYOZ}
\end{eqnarray}
In Eqs. \r{rohYOZ}-\r{dispPECYOZ}, $k_{z,(1)}$ and $k_{z,(2)}$ are
defined as in Sect. \ref{SubSectYOZFS}, $L$ is the thickness of the
slab, $\gamma _0 = \sqrt {k_y^2 - \omega ^2 \varepsilon _0 \mu _0
}$, $\gamma _h = \sqrt {k_y^2  - \omega ^2 \varepsilon _0 \mu _0
\varepsilon_h }$, and $k_y$ is the transverse wave number of the
incoming wave.
\end{widetext}
To illustrate the application of the formula, we have calculated the
phase of the reflection coefficient as a function of frequency for a
grounded metamaterial slab such that $a=L/10$ and different
incidence angles. The wires are assumed perfectly conducting and
thus the magnitude of $\rho$ is unity. The results obtained with Eq.
\r{rohYOZ} are depicted in Fig. \ref{phRevl} (solid lines)
superposed on data obtained using CST Microwave Studio (discrete
symbols) \cite{CST2008}. An excellent agreement is revealed
supporting the validity of the proposed homogenization methods.
It is interesting to note that the frequencies where the phase of
$\rho$ vanishes are nearly independent of the angle of incidence.
Thus, at such frequencies the grounded slab may mimic very closely
the behavior of an ideal perfect magnetic conductor
\cite{Sievenpiper}. Moreover, due to the large index of refraction
of the wire mesh such behavior may be obtained with a very
subwavelength slab. For example, the first resonance in the example
of Fig. \ref{phRevl} occurs for $L=0.02 \lambda_0$.

We have also calculated the dispersion characteristic $k_y = k_y
\left(\omega\right)$ of the guided modes supported by the grounded
slab. The dispersion characteristic is obtained by numerically
solving the equation $D(\omega, k_y)=0$ with respect to $k_y$, where
$D(\omega, k_y)$ is defined by Eq. \r{dispPECYOZ}. The calculated
results for a slab with fixed thickness $L$ and different values of
the lattice constant are depicted in Fig. \ref{sws} (dashed lines).
Consistent with the results reported in Ref. \cite{MarioEVL} and the
discussion of Sect. \ref{SubSectYOZFS}, it is seen that when the
density of wires is increased (i.e. $a$ is reduced) the guided mode
becomes more attached to the metamaterial slab. Unlike the case
where the slab stands in free-space the guided modes can only
propagate above a certain cut-off frequency. Despite that, due to
the anomalously high index of refraction of the crossed wire mesh,
the metamaterial slab thickness may be very subwavelength at the
onset of propagation of the fundamental mode \cite{MarioEVL,
EVLexp}. For example, for $a=L/10$ the thickness of the slab at the
cut-off frequency of the fundamental mode is as small as $L=0.02
\lambda_0$. Thus, the proposed structure may enable the realization
of ultra-subwavelength waveguides, as demonstrated experimentally in
\cite{EVLexp}.

It should be mentioned that, unlike a conventional dielectric slab,
the dispersion characteristic of a grounded crossed wires slab
cannot be obtained from the dispersion characteristic of a slab
standing in free-space. In fact, it is impossible to place a PEC
plane at the mid-plane of a crossed wires slab standing in
free-space without disturbing the fields, because the crossed wire
mesh is not transformed into itself after reflection with respect to
the $xoy$-plane.

\begin{figure}[th] \centering
\epsfig{file=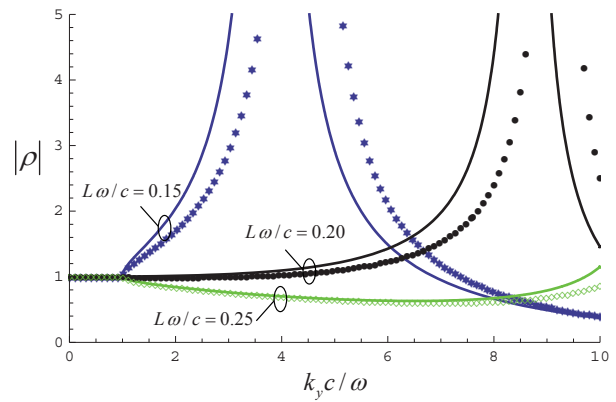, width=8.0cm} \caption{(Color online)
Amplitude of the reflection coefficient as a function of the
transverse component of the wave vector $k_y$, and different
frequencies of operation $\omega$. The lattice constant is $a =
L/10$ and the radius of the wires is $r_w=0.05a$, where $L$ is the
thickness of the grounded slab. Solid lines: Homogenization model.
Discrete symbols: full wave simulations.} \label{ampRevl}
\end{figure}

We have used CST Microwave Studio, in order to partially validate
the dispersion characteristics depicted Fig. \ref{sws}, obtained
using the homogenization model. Assuming plane wave incidence, we
have calculated the amplitude of the reflection coefficient as a
function of the transverse wave number of the incoming wave $k_y$,
for several fixed frequencies (Fig. \ref{ampRevl}).  For $k_y <
\omega/c$ the incoming wave is a propagating wave ($k_y=\omega/c
\sin \theta_i$), and, due to the conservation of energy, the
reflection coefficient amplitude is unity. On the other hand, for
$k_y > \omega/c$ the incoming wave is an evanescent mode, and thus
(since an evanescent wave does not carry power), $\left| \rho
\right| $ may be any nonnegative value, even larger than unity.
Indeed, it should be clear from Eq. \r{rohYOZ}, that the points
$k_y$ associated with the dispersion characteristic of the guided
modes correspond to the poles of the reflection coefficient. As seen
in Fig. \ref{ampRevl} there is good agreement between the response
to evanescent waves predicted by the homogenization model (solid
lines), and the actual response calculated using CST Microwave
Studio (discrete symbols). In particular, the position of the poles
is predicted with good accuracy, and thus these results partially
validate the dispersion characteristic depicted in Fig. \ref{sws}
associated with $a=L/10$ (dashed line).

\section{Propagation in the \emph{xoz} plane \label{SectXOZ}}

Here, we discuss the application of the proposed ABCs to the case
where the plane of incidence is the $xoz$-plane (Fig.
\ref{Isofreq}b). The geometry of the double-wire medium slab is as
explained in Sect. \ref{sectNum1}. It is assumed that the incoming
plane wave has parallel polarization (i.e. the electric field is in
the $xoz$-plane, whereas the incident magnetic field is along the
$y$-direction).

\subsection{Slab standing in free-space \label{SubSectXOZFS}}

Considering the geometry of the system under-study and the
polarization of the incoming wave, it is obvious that the magnetic
field in all space has a single Cartesian component, $H_y$. As in
Sect. \ref{SubSectYOZFS}, it is assumed that inside the metamaterial
slab the field is written in terms of a superposition of plane
waves. The propagation constants along $z$ of the plane waves can be
calculated by solving the dispersion characteristic \r{dispEqXOZ}
with respect to $k_z$, for a fixed frequency $\omega$, and a fixed
transverse wave number of the incident wave $k_x = \omega/c \sin
\theta_i$. It can be verified that the dispersion characteristic is
equivalent to a polynomial equation of degree three in the variable
$k_z^2$ \cite{XWiresNegRef}. Thus, the metamaterial slab supports
three independent plane waves with magnetic field along the
$y$-direction. Clearly, such property is a consequence of the
nonlocal effects, since in a local material there cannot exist two
distinct plane waves (i.e. associated with different $k_z^2$) with
the same magnetic field orientation. The propagation constants along
$z$ of the plane waves in the bulk metamaterial are denoted by
$k_{z,(1)}$, $k_{z,(2)}$, and $k_{z,(3)}$. Thus, the magnetic field
inside the metamaterial slab ($-L<z<0$) can be written as (the
$x$-variation of the fields is suppressed),
\begin{eqnarray}
H_y  &=& A_1^ +  e^{ + i\,k_{z,(1)} z }  + A_1^ -  e^{ - i\,k_{z,(1)} z } + A_2^ +  e^{ + i\,k_{z,(2)} z }  + \nonumber \\
&& A_2^ -  e^{ -i\,k_{z,(2)} z }  + A_3^ +  e^{ + i\,k_{z,(3)} z } +
A_3^ - e^{ - i\,k_{z,(3)} z }, \nonumber \\
&& -L<z<0 \l{HyfieldXOZ}
\end{eqnarray}
where $A_i^{\pm}$ are the complex amplitudes of the excited waves.
For frequencies much lower than the plasma frequency of the wire
medium, $\omega/c \ll \beta_p /\sqrt{\varepsilon_h}$, it can be
verified that in the PEC case and for a propagating incoming wave,
only a single propagation constant, let's say $k_{z,(1)}$, is real
valued. The corresponding plane wave is associated with the
hyperbolic isofrequency contours depicted in Fig. \ref{Isofreq}b.
The other two propagation constants, $k_{z,(2)}$ and $k_{z,(3)}$ are
pure imaginary, and are associated with evanescent modes.

Assuming that the incoming plane wave propagates in the semi-space
$z>0$, the magnetic field in the air regions verifies,
\begin{eqnarray}
 H_y  &=& H_y^{inc} \left( {e^{\gamma _0 z}  - \rho \, e^{ - \gamma _0 z} } \right),\quad z > 0 \nonumber \\
 H_y  &=& H_y^{inc} Te^{\gamma _0 z} ,\quad z <  - L
 \l{Hyair}
\end{eqnarray}
where $H_y^{inc}$ is the complex amplitude of the incident magnetic
field, $\gamma _0  =  - i\sqrt {\omega ^2 \varepsilon _0 \mu _0 -
k_x^2 }  = \sqrt {k_x^2  - \omega ^2 \varepsilon _0 \mu _0 }$ is the
propagation constant of the incoming wave along the $z$-direction,
and $\rho$ and $T$ are the reflection and transmission coefficients,
respectively.

The electric field inside the metamaterial slab can be easily
obtained from Eq. \r{HyfieldXOZ}, by noting that for each individual
plane wave of the type ${\bf{H}} = H_{0} e^{i{\kern 1pt} k_x x}
e^{i\,k_z z} {\bf{\hat u}}_y$, the corresponding electric field is
${\bf{E}} = H_{0} \, \dyad{\varepsilon}^{ - 1} {\bf{.}}\left(
{\frac{{k_z }}{{\omega \varepsilon _0 }}{\bf{\hat u}}_x  -
\frac{{k_x }}{{\omega \varepsilon _0 }}{\bf{\hat u}}_z } \right)
e^{i{\kern 1pt} k_x x} e^{i\,k_z z}$, where $\dyad{\varepsilon}^{ -
1}\left( {\omega ,{\bf{k}}} \right)$ is the inverse of the
dielectric function of the crossed wire mesh, defined by Eq.
\r{epsWM} with $N=2$. The averaged macroscopic current $
{\bf{J}}_{d,{\rm{av}}}$ inside the metamaterial is finally obtained
by replacing the electric field in formula \r{Jdav_fnE}, taking into
account that for the present geometry ${\bf{k}}_{||}  = k_x
{\bf{\hat u}}_x$. The formulas for $\bf{E}$ and $
{\bf{J}}_{d,{\rm{av}}}$ are too long to show here, and thus are
omitted.

As in section \ref{SectYOZ}, the reflection and transmission
coefficients can be calculated by imposing the continuity of the
tangential electromagnetic fields ($H_y$ and $E_x$ components of the
fields), and the two additional boundary conditions \r{ABCair}
($N=2$). Unlike in section \ref{SectYOZ}, the two ABCs are not
degenerate for the present geometry, and yield two independent
equations. This property is consistent with the fact that the wire
medium supports two additional waves for the present configuration,
as is manifest from Eq. \r{HyfieldXOZ}, and thus two ABCs are
required to remove the two additional degrees of freedom.
\begin{figure}[th] \centering
\epsfig{file=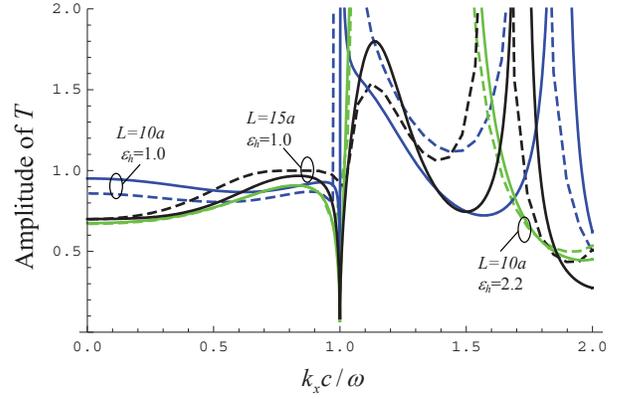, width=8.0cm} \caption{(Color online)
Amplitude of the transmission coefficient as a function of
normalized $k_x$ for a fixed frequency and different $L$ and
$\varepsilon_h$. The lattice constant is such that $\omega a/c
=0.6$, the radius of the wires is $r_w=0.05a$. Solid lines:
Homogenization model. Dashed lines: full wave simulations.}
\label{ampTkxhyp}
\end{figure}
\begin{figure}[th] \centering
\epsfig{file=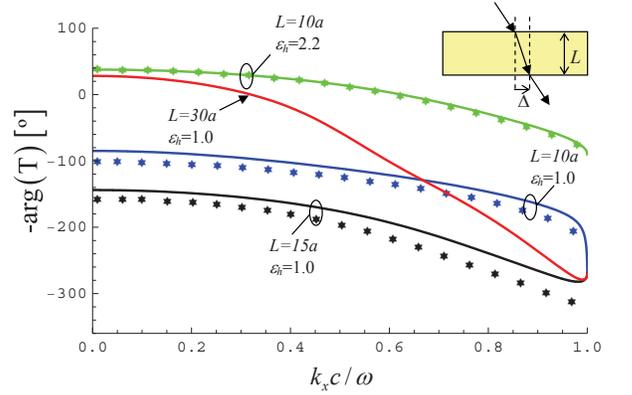, width=8.0cm} \caption{(Color online)
Phase of the transmission coefficient as a function of normalized
$k_x$ for a fixed frequency and different $L$ and $\varepsilon_h$.
The lattice constant is such that $\omega a/c =0.6$, the radius of
the wires is $r_w=0.05a$. Solid lines: Homogenization model. Dashed
lines: full wave simulations.} \label{phTkxhyp}
\end{figure}

We have applied the proposed homogenization procedure to
characterize the response of a wire medium slab under plane wave
incidence. In Fig. \ref{ampTkxhyp} we depict the amplitude of the
transmission coefficient, for a fixed frequency and lattice constant
$a$, with $\omega a/c=0.6$, and for different values of the slab
thickness $L$, as a function of the transverse wave number $k_x$ of
the plane wave. As mentioned before $k_x = \omega/c \sin \theta_i$
for a propagating plane wave, whereas $k_x > \omega/c$ when the
incoming wave is evanescent. For simplicity the wires are assumed
perfectly conducting. The results of Fig. \ref{ampTkxhyp} reveal a
good agreement between the homogenization results (solid lines) and
the full wave simulations obtained with CST Microwave Studio (dashed
lines), both for the propagating spectrum ($k_x < \omega/c$), and
for the evanescent spectrum ($k_x > \omega/c$). The homogenization
results are especially accurate in the example where the
permittivity of the host material is $\varepsilon_h=2.2$. It is seen
that the transmission coefficient may have several poles which
indicate the presence of guided modes.

As reported in Ref. \cite{XWiresNegRef}, as a consequence of the
hyperbolic shaped isofrequency contours associated with the
propagating mode (Fig. \ref{Isofreq}b), the Poynting vector suffers
negative refraction at the interface between a crossed wire mesh and
air. Therefore, unlike in a conventional dielectric slab, the
incident beam suffers a negative lateral shift $\Delta$ as it
propagates through the wire medium slab \cite{XWiresNegRef} (the
inset in Fig. \ref{phTkxhyp} represents the case in which $\Delta$
is positive). As demonstrated in Ref. \cite{XWiresNegRef}, the
spatial shift $\Delta$ can be related to the phase $\phi  =  - \arg
T $ of the transmission coefficient ($T = \left| T \right|e^{ -
i{\kern 1pt} \phi }$), by the simple formula $\Delta  = d\phi
/dk_x$. Thus, the spatial shift is proportional to the slope of the
phase of $T$. Hence, when $\phi$ is an increasing function of $k_x$
the lateral shift is positive and the beam is positively refracted
(this is what happens in a conventional dielectric slab). On the
other hand, when $\phi$ is a decreasing function of $k_x$ the
lateral shift is negative, which indicates that the beam is
negatively refracted \cite{XWiresNegRef}.

In order to show that our homogenization model predicts the
emergence of negative refraction, we have calculated the phase of
the transmission coefficient as a function of $k_x = \omega/c \sin
\theta_i$, for the normalized frequency $\omega a/c =0.6$ ($\omega$
and $a$ are fixed). The obtained results are represented in Fig.
\ref{phTkxhyp} for different thicknesses of the metamaterial slab.
The discrete symbols were calculated using CST Microwave Studio. It
is seen in Fig. \ref{phTkxhyp} that, consistent with the results
reported in \cite{XWiresNegRef}, the phase $\phi$ is indeed a
decreasing function of $k_x$. In particular, it is seen that for
thicker slabs the slope of $\phi$ becomes more negative, which
indicates, as could be expected, that the lateral spatial shift
becomes larger. Similarly, the slope also increases with the angle
of incidence (i.e. with $k_x$), consistent with the fact that for
larger values of $\theta_i$ the beam is more refracted at the
interface \cite{XWiresNegRef}. As demonstrated in Ref.
\cite{XWiresNegRef}, the phenomenon of negative refraction is very
broadband.

\subsection{Grounded Slab}

\begin{figure}[th] \centering
\epsfig{file=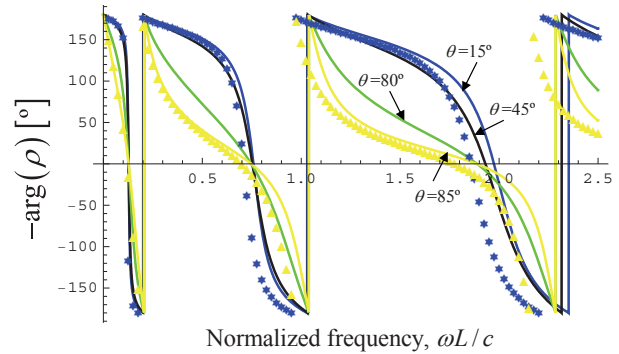, width=8.0cm} \caption{(Color
online) Phase of the reflection coefficient as a function of
normalized the normalized frequency for different angles of
incidence. The lattice constant is $a = L/10$ and the radius of the
wires is $r_w=0.05a$, where $L$ is the thickness of the grounded
slab. Solid lines: Homogenization model. Star and triangle shaped
symbols: full wave simulations for $\theta=15^o$ and $\theta=85^o$,
respectively.} \label{phRhyp}
\end{figure}

For the sake of completeness, we have also studied the application
of the ABCs to the case where the bottom face of the metamaterial
slab is grounded ($z=-L$). As in Sect. \ref{SubSectXOZFS}, the
magnetic field inside the wire medium is written as in Eq.
\r{HyfieldXOZ}. The magnetic field in the air region ($z>0$) is
defined in the same manner as in Eq. \r{Hyair}. For this
configuration, the homogenization procedure is similar to that
described in Sect. \ref{SubSectXOZFS}, except that at the ground
plane it must be enforced that $E_x = 0$ and the two ABCs
\r{ABCmetal} ($N=2$). We have applied such analytical formalism to
characterize the response of a grounded metamaterial slab with a
fixed thickness $L$ and lattice constant $a=L/10$, as a function of
frequency. In Fig. \ref{phRhyp} we depict the calculated reflection
coefficient phase (solid lines) for different angles of incidence.
The host material is air and the wires are perfectly conducting.
Similar to the results of Sect. \ref{SubSectYOZGround}, it is seen
that at certain frequencies, nearly independent of the angle of
incidence, the metamaterial may behave as a high impedance ground
plane, and mimic very closely the behavior of a perfect magnetic
conductor. It should be noted that in the present configuration the
incoming wave is TM-polarized, whereas in the Sect.
\ref{SubSectYOZGround} the wave is TE-polarized. Thus, the high
impedance property is independent of the polarization. The
homogenization results concur well with the full wave simulations
computed with CST Microwave Studio (discrete symbols), which further
validates our homogenization theory.

\section{Conclusion \label{SectConclusion}}

We extended our previous work on additional boundary conditions
\cite{MarioABC, ABCtilted} to the case of wire media formed by $N$
different nonconnected components ($N=1,2,3$). Using simple physical
arguments, it was demonstrated that in general $N$ different ABCs
must be considered at an interface. These ABCs were derived for the
cases where the wire medium is either adjacent to a dielectric or to
a conducting material. It was proven that in the absence of loss the
proposed ABCs ensure the conservation of the power flow, and in
particular we have derived a general formula for the Poynting vector
in the wire medium. We have illustrated the application of the
proposed ABCs when the metamaterial is formed by a double array of
metallic wires ($N=2$). It was demonstrated that the proposed
homogenization concepts enable the accurate numerical modeling of
the considered nonlocal materials in scattering problems (for both
propagating and evanescent incident waves) and in propagation
problems (calculation of the guided modes). In particular, we have
highlighted the anomalous physical properties of nonconnected wire
media, and demonstrated that these materials may enable the
realization of very compact devices and the emergence of negative
refraction.

\begin{acknowledgments}
This work is supported in part by Funda\c{c}\~ao para a Ci\^encia e
a Tecnologia under project PDTC/EEA-TEL/71819/2006.
\end{acknowledgments}

\appendix
\section{\label{AppTABulk}}

In this Appendix we demonstrate that in nonconnected wire media the
bulk macroscopic fields may be identified to a good approximation
with the TA-fields, provided none of the unit vectors ${\bf{\hat
u}}_n$ lies in the transverse ($xoy$) plane. This result is a
generalization to double and triple wire media of a similar property
characteristic of arrays of parallel wires \cite{ABCtilted}.

To this end, consider the unbounded periodic material and suppose
first that the microscopic fields $\left( {{\bf{e}},{\bf{b}}}
\right)$ have the Floquet-Bloch property, i.e. $ \left(
{{\bf{e}},{\bf{b}}} \right)e^{ - i{\bf{k}}{\bf{.r}}}$ has the same
translational symmetry as the lattice, where $\bf{k}$ is the
associated wave vector. Following Refs. \cite{Silv_MTT_3DWires,
Silv_TAfield} the bulk electric field is defined by ${\bf{E}} =
{\bf{E}}_{{\rm{av}}} e^{i{\bf{k}}{\bf{.r}}}$ with (using the time
convention $e^{-i \omega t}$): \e {\bf{E}}_{{\rm{av}}}  =
\frac{1}{{{\rm{V}}_{{\rm{cell}}} }}\int\limits_\Omega
{{\bf{e(r)}}e^{-i{\kern 1pt} {\bf{k}}{\bf{.r}}} d^3 {\bf{r}}} \f The
macroscopic induction field ${\bf{B}}_{{\rm{av}}}$ is defined
similarly. In the above, $\Omega$ represents the unit cell of the
nonconnected wire medium, and ${\rm{V}}_{\rm{cell}} = a^3$ is its
volume. The bulk macroscopic fields verify the system
\cite{Silv_MTT_3DWires, Silv_TAfield}:
\begin{eqnarray}
i{\bf{k}} \times \frac{{{\bf{B}}_{{\rm{av}}} }}{{\mu _0 }} &=&
-i\omega \varepsilon _0 \varepsilon _h {\bf{E}}_{{\rm{av}}} -i\omega
{\bf{P}}_{{\rm{av}}} \nonumber \\
  i{\bf{k}} \times {\bf{E}}_{{\rm{av}}}  &=& i\omega \,{\bf{B}}_{{\rm{av}}}
\l{Max1}
\end{eqnarray}
where the generalized polarization vector, ${\bf{P}}_{{\rm{av}}}$,
is given by, \e {\bf{P}}_{{\rm{av}}} = \frac{{\,1}}{{ -i
\omega{\rm{V}}_{{\rm{cell}}} }} \int\limits_{\partial D }
{{\bf{J}}_c{\bf{(r)}} } e^{ -i\,{\bf{k}}{\bf{.r}}} ds \f  and
${\bf{J}}_c$ is the density of current over the surface $\partial D$
of the metallic wires enclosed in the unit cell. The dielectric
function of the bulk material (defined by Eq. \r{epsWM}) is such
that $\dyad{ \varepsilon } \left( {\omega ,{\bf{k}}}
\right){\bf{.E}}_{{\rm{av}}} = \varepsilon _0 \varepsilon _h
{\bf{E}}_{{\rm{av}}}  + {\bf{P}}_{{\rm{av}}}$
\cite{Silv_MTT_3DWires}.

The objective here is to relate the bulk medium fields with the
TA-fields defined by \r{TAdef}. Notice that unlike the bulk medium
fields, the definition of the TA-fields depends on the considered
transverse plane, or equivalently, depends on the orientation of the
wires relative to the transverse plane (which is assumed in this
work to be the $xoy$ plane). We suppose that the transverse unit
cell $\Omega _T$ may be related to the unit cell of the bulk
material as, $\Omega = \Omega _T \times \left[ { - a_z /2,\;a_z /2}
\right]$ for some $a_z$ which may depend on the orientation of the
wires relative to the transverse plane. From Eq. (26) of
Ref.\cite{Silv_TAfield} it is known that the generalized
polarization vector may be written as a function of the TA-density
of current as,
\begin{eqnarray} {\bf{P}}_{{\rm{av}}}  =  \frac{1}{{-i \omega a_z
}}\int\limits_{ -a_z/2}^{a_z /2} {{\bf{J}}_{d,{\rm{av}}} \left( z
\right)e^{ -i{\kern 1pt} k_z z} dz} \l{Pg_jdav}\end{eqnarray} where
$k_z$ is the $z$ component of the wave vector. But, as demonstrated
in section \ref{ABC}, within the thin wire approximation, the
TA-current is expressed in terms of the microscopic currents as in
Eq. \r{Jdavsimp}. Moreover, since we assume here that the
microscopic fields verify the Floquet-Bloch condition (along three
independent directions of space) it is clear that the microscopic
currents must be such that $I_n(z) = I_n e^{i k_z z}$, for some
$I_n$. Taking this property into account, and using Eq.
\r{Jdavsimp}, it is immediate that Eq. \r{Pg_jdav} implies that:
\e{\bf{J}}_{d,{\rm{av}}} \left( z \right) =  - i\omega
{\bf{P}}_{{\rm{av}}} {\kern 1pt} e^{i{\kern 1pt} k_z
z}\l{JdavPgWM}\f Therefore, within the thin wire approximation, and
independent of the orientation of the wires relative to the
interface, the TA-current associated with a Floquet mode of the
unbounded material can be written in terms of the polarization
vector of the bulk material as in Eq. \r{JdavPgWM}. Such result
readily implies that the TA-fields must verify:
\e{\bf{E}}_{{\rm{av,T}}} \left( z \right) = {\bf{E}}_{{\rm{av}}}
{\kern 1pt} e^{i{\kern 1pt} k_z z} ,\quad \;{\bf{B}}_{{\rm{av,T}}}
\left( z \right) = {\bf{B}}_{{\rm{av}}} {\kern 1pt} e^{i{\kern 1pt}
k_z z} \l{TAbulk}\f Indeed, in the considered scenario, the
TA-fields are univocally determined by the ($k_z$-Floquet) solution
of the differential system \r{TAfields} with
${\bf{J}}_{d,{\rm{av}}}$ given by Eq. \r{JdavPgWM}. But since the
bulk medium fields verify Eqs. \r{Max1}, and noting that the wave
vector can be decomposed as ${\bf{k}} = {\bf{k}}_{||} + k_z
{\bf{\hat u}}_z $, being ${\bf{k}}_{||}$ the projection of the wave
vector into the transverse plane, it readily follows that, indeed,
the solution of the differential system \r{TAfields} is given by
\r{TAbulk}. This demonstrates that the TA-fields can be identified
with the bulk medium fields, as we wanted to show. It should also be
clear that Eq. \r{TAbulk} can be immediately generalized to the case
where the microscopic fields are a superposition of several
Floquet-Bloch modes (associated with different wave vectors). In
such case, the TA-fields are obviously given by a superposition of
plane waves (being each plane wave associated with a different
Floquet mode as in Eq. \r{TAbulk}).

As a final remark, we note that the result \r{TAbulk} is only valid
provided there are no wires parallel (or quasi-parallel) to the
interface, because, as noted in section \ref{ABC}, Eq. \r{Jdav}
becomes singular in such circumstances. In fact, if some of wires
are parallel to the interface, it was demonstrated in
Ref.\cite{Silv_TAfield} that the TA-fields cannot be really
identified with the bulk macroscopic fields. The reader is referred
to Ref.\cite{Silv_TAfield} for more details about the methods that
can be used to study such configurations.

\section{\label{AppSzPW}}

In this Appendix we demonstrate that for the superposition of plane
waves \r{PWs}, $S_z$ defined by Eq. \r{Sz} may be written as in Eq.
\r{SzPW}.

To begin with, we note that the polarization vector associated with
the superposition of plane waves is: \e{\bf{P}}\left( {\bf{r}}
\right) = \varepsilon _0 \sum\limits_l {\left( {\overline{\overline
\varepsilon } \left( {\omega ,{\bf{k}}_l } \right) - \varepsilon _h
\overline{\overline {\bf{I}}} } \right){\bf{.E}}_l e^{ + i{\kern
1pt} {\bf{k}}_l {\bf{.r}}} } \f where $\dyad{\varepsilon}$ is the
dielectric function of the wire medium. In particular, the
projection of the ${\bf{P}}$ onto the direction ${\bf{\hat u}}_n$
(parallel to one of the wire arrays) verifies, \e P_n  = \varepsilon
_0 \sum\limits_l {\left( {\varepsilon _{n,n} \left( {\omega
,{\bf{k}}_l } \right) - \varepsilon _h } \right){\bf{\hat u}}_n
{\bf{.E}}_l e^{ + i{\kern 1pt} {\bf{k}}_l {\bf{.r}}} } \l{ApBaux1}\f
and,
\begin{eqnarray}
&\left( {i{\bf{k}}_{||}  + \frac{d}{{dz}}{\bf{\hat u}}_z }
\right).{\bf{\hat u}}_n P_n&  = \varepsilon _0 \sum\limits_l {\left(
{i{\kern 1pt} {\bf{k}}_l .{\bf{\hat u}}_n } \right)}  \times
\nonumber \\
&&\left( {\varepsilon _{n,n} \left( {\omega ,{\bf{k}}_l } \right) -
\varepsilon _h } \right){\bf{\hat u}}_n {\bf{.E}}_l e^{ + i{\kern
1pt} {\bf{k}}_l {\bf{.r}}}\nonumber \\ \l{ApBaux2}
\end{eqnarray}
On the other hand, explicit calculations show that: \e \left(
{\varepsilon _{n,n}  - \varepsilon _h } \right)^{ - 1}
\frac{{\partial \varepsilon _{n,n} }}{{\partial k_z }} =
\frac{1}{{\varepsilon _h \beta _p^2 }}\left( {\varepsilon _{n,n}  -
\varepsilon _h } \right)\left( { - 2{\bf{k}}{\bf{.\hat u}}_n }
\right)\left( {{\bf{\hat u}}_z {\bf{.\hat u}}_n } \right)
\l{ApBaux3}\f Substituting Eqs. \r{ApBaux1}-\r{ApBaux3} and Eq.
\r{PWs} into Eq. \r{Sz}, it is found that:
\begin{widetext}
\e S_z  = S_z^0  - \frac{\omega }{4}\varepsilon _0 \sum\limits_{l,m}
{{\mathop{\rm Re}\nolimits} } \left\{ {\sum\limits_{n = 1}^N {\left(
{\varepsilon _{n,n} \left( {\omega ,{\bf{k}}_l } \right) -
\varepsilon _h } \right)^{ - 1} \left( {\varepsilon _{n,n} \left(
{\omega ,{\bf{k}}_m } \right) - \varepsilon _h } \right)^*
{\bf{E}}_m^* {\bf{.\hat u}}_n \frac{{\partial \varepsilon _{n,n}
}}{{\partial k_z }}\left( {\omega ,{\bf{k}}_l } \right){\bf{\hat
u}}_n {\bf{.E}}_l } e^{i\left( {{\bf{k}}_l  - {\bf{k}}_m^* }
\right){\bf{.r}}} } \right\} \l{ApBSzaux}\f \end{widetext} This
result and the definition of $S_z^0$ \r{Sz0def}, show that $S_z$ may
be written as a linear combination of exponentials of the type
$e^{i\left( {{\bf{k}}_l  - {\bf{k}}_m^* } \right){\bf{.r}}}$, where
$l$ and $m$ index a generic plane wave from the considered set.
However, from Eq. \r{conservSz}, we know that in the lossless case
$S_z$ must be a constant, i.e. independent of $\bf{r}$. This means
that in Eq. \r{ApBSzaux} the coefficients associated with the
exponential $e^{i\left( {{\bf{k}}_l - {\bf{k}}_m^* }
\right){\bf{.r}}}$ necessarily vanish when ${\bf{k}}_l  \ne
{\bf{k}}_m^*$. Hence, using the properties $\varepsilon _{n,n}
\left( {\omega ,{\bf{k}}_m } \right)^*  = \varepsilon _{n,n} \left(
{\omega ,{\bf{k}}_m^* } \right)$ and $\frac{{\partial
\overline{\overline \varepsilon } }}{{\partial k_z }} =
\sum\limits_{n = 1}^N {\frac{{\partial \varepsilon _{n,n}
}}{{\partial k_z }}{\bf{\hat u}}_n {\bf{\hat u}}_n }$, we finally
conclude that $S_z$ may be written as in Eq. \r{SzPW}.

\bibliography{refs}
\end{document}